\documentclass[letterpaper, 10 pt, conference]{ieeeconf}  

\IEEEoverridecommandlockouts                              

\usepackage{amsmath}
\usepackage{booktabs}
\usepackage{tabularx}
\usepackage{graphicx}
\usepackage{xcolor} 
\usepackage{multirow}
\usepackage{cite}
\let\labelindent\relax
\usepackage{enumitem} 
\usepackage{adjustbox}

\usepackage{url}

\title{\LARGE \bf
Heterogeneous Mixed Traffic Control and Coordination 
}

\author{Iftekharul Islam$^{1}$, Weizi Li$^{1}$, Xuan Wang$^{2}$, Shuai Li$^{3}$, and Kevin Heaslip$^{4}$ 
\thanks{$^{1}$Iftekharul Islam and Weizi Li are with Min H. Kao Department of Electrical Engineering and Computer Science at University of Tennessee, Knoxville, TN, USA {\tt\small mislam73@vols.utk.edu, weizili@utk.edu}}%
\thanks{$^{2}$Xuan Wang is with Department of Electrical and Computer Engineering at George Mason University, Fairfax, VA, USA {\tt\small xwang64@gmu.edu}}%
\thanks{$^{3}$Shuai Li is with Department of Civil and Coastal Engineering at University of Florida, Gainesville, FL, USA {\tt\small shuai.li@ufl.edu}}%
\thanks{$^{4}$Kevin Heaslip is with Department of Civil and Environmental Engineering at University of Tennessee, Knoxville, TN, USA {\tt\small kheaslip@utk.edu}}
}

\begin{document}

\maketitle
\thispagestyle{empty}
\pagestyle{empty}

\begin{abstract}

Urban intersections with diverse vehicle types, from small cars to large semi-trailers, pose significant challenges for traffic control. This study explores how robot vehicles (RVs) can enhance heterogeneous traffic flow, particularly at unsignalized intersections where traditional methods fail during power outages. Using reinforcement learning (RL) and real-world data, we simulate mixed traffic at complex intersections with RV penetration rates ranging from 10\% to 90\%. Results show that average waiting times drop by up to 86\% and 91\% compared to signalized and unsignalized intersections, respectively. We observe a ``rarity advantage,'' where less frequent vehicles benefit the most (up to 87\%).  
Although CO$_2$ emissions and fuel consumption increase with RV penetration, they remain well below those of traditional signalized traffic.
Decreased space headways also indicate more efficient road usage. These findings highlight RVs' potential to improve traffic efficiency and reduce environmental impact in complex, heterogeneous settings.

\end{abstract}


\section{INTRODUCTION}
\label{introduction}
%

Urban roadways are dominated by heterogeneous traffic, characterized by a mix of vehicles such as passenger cars, trucks, vans, and more. The complexities arising from the varying sizes, speeds, and maneuverability of these vehicle types significantly affect overall road performance~\cite{kong2021modeling}. For example, larger vehicles like trucks require more space and tend to cause delays, while smaller cars navigate more easily~\cite{mohan2020capacity, mohan2021investigating}. These differences often disrupt traffic flow, as slower vehicles interfere with the movement of faster ones, increasing instability and complicating traffic management~\cite{sun2021modeling}. As a result, traffic congestion intensifies, causing extended commute times and significant economic impacts. In the U.S. alone, drivers lose an average of 100 hours annually to traffic congestion, resulting in economic losses of \$1,400 per person~\cite{inrix2019}.

Intersections play a crucial role in urban traffic systems where vehicles converge and disperse. Traditionally, traffic lights are used to manage these complex interactions. However, intersections can turn into congestion points, especially when such control methods fail or are absent. Natural disasters, large-scale power failures, or system malfunctions can disable traffic signals, converting typically controlled intersections into unregulated crossings. Additionally, during temporary road configuration due to construction, intersections may lack signalization. These various situations pose significant traffic management challenges for urban areas, forcing them to navigate complex traffic conditions without traffic lights for extended periods~\cite{outage, outage2}. 


The emergence of robot vehicles (RVs) with varying degrees of self-driving capabilities has the potential to transform urban traffic management. These vehicles can be leveraged to manage nearby human-driven vehicles (HVs) so that the \emph{mixed traffic}, consisting of both RVs and HVs, can be optimized for various objectives such as improving efficiency or reducing emissions~\cite{rios2016survey, li2017platoon, villarreal2023mixed, yue2023revolution, villarreal2024analyzing, di2021survey}. 
When selecting control strategies for mixed traffic, reinforcement learning (RL) has demonstrated significant potential due to its model-free characteristics~\cite{di2021survey, vinitsky2018benchmarks, wang2023learning}. RL is particularly well-suited for managing multi-agent systems at complex and dynamic environments, e.g., traffic at unsignalized intersections, because of its flexibility to adapt and learn effective strategies~\cite{sutton2018reinforcement}. 
Unlike traditional rule-based systems, RL can potentially discover novel solutions that might not be apparent to human traffic engineers~\cite{han2023leveraging}. This combination of RVs and RL presents a promising approach for addressing the challenges of mixed traffic control and coordination. 

Despite the aforementioned advancements, a significant gap exists between current research of mixed traffic and the complexities of real-world traffic: nearly all existing studies assume \emph{homogeneous traffic}.   
Real-world traffic, on the other hand, features a variety of vehicle types, i.e., \emph{heterogeneous traffic},  equipped with a wide range of kinematic and dynamic parameters. 
These factors are essential for effective traffic control and management~\cite{kong2021modeling,mohan2020capacity,mohan2021investigating,sun2021modeling}.
The limitations of homogeneous traffic become more pronounced when factoring in the complexity of test environments and varying RV penetration rates, raising several important research questions: 1) Is it feasible to implement heterogeneous mixed traffic control in \emph{complex environments}? 2) What are the performance implications for entire traffic and different vehicle types in terms of \emph{efficiency}, \emph{environmental impact}? and 3) How do \emph{varying RV penetration rates} affect performance? 

\noindent Addressing these questions will significantly enhance the applicability of mixed traffic control and coordination, particularly in complex environments like unsignalized intersections where traditional control methods may be absent.



In this study, we aim to fill the critical gap by demonstrating the feasibility of heterogeneous mixed traffic control and coordination in complex environments. 
We also conduct comprehensive experiments to evaluate efficiency and environmental impact of the heterogeneous traffic.
Specifically, our contributions are the following.
\begin{itemize}[leftmargin=*]
    \item Our approach incorporates multiple vehicle types, including passenger cars, pickups, vans, semi-trailers, and trucks, extracted from real-world traffic data. 
    \item Our method is implemented and tested across multiple complex, real-world intersection layouts.
    \item Our experiments are grounded in vehicle type distributions that mirror actual urban traffic composition.
    \item Our detailed analysis examines various RV penetration rates to understand the impact of gradual automation in mixed traffic environments.
\end{itemize}

\noindent To the best of our knowledge, this is the first study investigating the RL-based control and coordination of heterogeneous mixed traffic in complex environments. The findings and insights extend the frontier of emerging mixed traffic research.

\section{RELATED WORK}
\label{related_work}


The diversity in vehicle characteristics within heterogeneous traffic poses numerous challenges for traffic management. Studies show that the composition of traffic significantly impacts overall road performance, with larger and slower vehicles often exacerbating congestion and causing delays~\cite{kong2021modeling, ngoduy2015effect}. Sun et al.~\cite{sun2021modeling} demonstrate that considering traffic heterogeneity, particularly leader-follower compositions and driving styles, is crucial for accurately modeling traffic dynamics, as neglecting these factors can lead to substantial estimation errors.

At intersections, these challenges are more pronounced. Mohan and Chandra~\cite{mohan2021investigating} reveal that larger vehicles like trucks and buses require longer gaps to navigate intersections safely, which reduces overall intersection capacity and increases delays. Furthermore, safety evaluations and conflicting vehicle speeds indicate that interactions between diverse vehicle types greatly affect the likelihood of conflicts~\cite{hasain2022safety}. Suriyarachchi et al.~\cite{suriyarachchi2024gameopt+} propose a game-theoretic framework to manage heterogeneous traffic at intersections. Their approach optimizes the sequence of vehicle entries at unsignalized intersections to reduce fuel consumption. However, game-theoretic methods may face scalability issues in more unpredictable traffic scenarios~\cite{liu2023game}.

Several studies explore the management of traffic at unsignalized intersections, particularly for connected and autonomous vehicles (CAVs). Early work by Dresner and Stone~\cite{dresner2008multiagent} proposes a multi-agent intersection control system where CAVs reserve space-time slots using first-come, first-serve (FCFS). Jin et al.~\cite{jin2013platoon} expand on this pioneer work by applying FCFS scheduling to CAV platoons, while Chen et al.~\cite{chen2022improved} propose a controllable gap strategy that adjusts the time gap between vehicles based on their speed and conflict relationships to prevent collisions. Researchers also examine decentralized approaches, such as the consensus-based control method for managing the trajectories of CAVs by Mirheli et al.~\cite{mirheli2019consensus} and Malikopoulos et al.'s work~\cite{malikopoulos2018decentralized} for optimizing the energy consumption. These methods typically assume a fully autonomous environment.


Recently, studies show the effectiveness of RL in managing and optimizing mixed traffic across various scenarios. Wang et al.~\cite{wang2023learning,Wang2024Privacy} demonstrate that decentralized multi-agent RL can significantly reduce waiting times at complex unsignalized intersections with high RV penetration. Villarreal et al.~\cite{villarreal2023mixed} explore image-based observations as a substitute for precise traffic data in RL-based control. 
Poudel et al.~\cite{Poudel2024EnduRL,poudel2024carl} propose an RL framework that integrates real-world driving profiles to improve safety, stability, and efficiency. 
Yan and Wu~\cite{yan2021reinforcement} introduce a model-free multi-agent RL approach for controlling mixed traffic at unsignalized intersections. Peng et al.~\cite{peng2021connected} apply RL to boost efficiency at such intersections using CAVs. 
Shi et al.~\cite{shi2022control} show that RL can reduce collisions and improve flow in urban unsignalized settings over traditional control methods.

Despite significant progress, most existing studies face two major limitations: 1) homogeneous traffic is assumed, and 2) the test environment is simplified (either by limiting the number of conflicting vehicles or using idealized intersection topologies).
This lack of real-world complexity hinders the  applicability of the proposed techniques.
We aim to fill the gap by demonstrating the effectiveness of RL for managing heterogeneous mixed traffic in complex  environments.

\section{METHODOLOGY}
\label{methodology}




\subsection{RL-based Heterogeneous Traffic Control}

We formulate the problem of managing heterogeneous mixed traffic at unsignalized intersections as a Partially Observable Markov Decision Process (POMDP), defined as the collection \((\mathcal{X}, \mathcal{U}, \mathcal{P}, \mathcal{R}, \mathcal{Y}, \mathcal{V}, \gamma)\). Here, \(\mathcal{X}\) represents the set of all possible states; \(\mathcal{U}\) denotes the set of available actions; \(\mathcal{P}({x}'|{x}, u)\) describes the probability of transitioning to a new state \({x}'\) given the present state \({x}\) and action \(u\); and \(\mathcal{R}({x}, u)\) defines the reward function that provides feedback based on the state-action pair. The observation space \(\mathcal{Y}\) captures the partial view of the environment each RV receives, with \(\mathcal{V}(y|{x})\) representing the probability of receiving an observation \(y\) from the state \({x}\). The discount factor $\gamma \in [0,1)$ determines the weighting of future rewards.
At each step \(t\), an RV chooses an action \(u_t \in \mathcal{U}\) using the policy \(\pi_\theta(u_t|{x}_t)\). The environment transitions to the next state \({x}_{t+1}\), and the RV receives a reward \(r_t\). 
The goal is to maximize the total discounted reward: \(\mathcal{R}_t = \sum_{i=t}^{T} \gamma^{i-t} r_i\)~\cite{wang2023learning}.






RVs decide between two possible actions: Stop and Go. Stop halts the RV before entering the intersection, while Go allows it to proceed.
This high-level, binary action space allows the policy to focus on strategic coordination, while the precise acceleration and deceleration are handled by low-level controllers described in Sec.~\ref{veh_behavior}.
The observation \(y_t\) of an RV at time \(t\) encompasses traffic conditions at the intersection: $y_t = \oplus_{k \in K} \langle l_k, w_k \rangle \oplus_{k \in K} \langle o_k \rangle \oplus \langle d_{int} \rangle$,
\noindent where $K$ is the set of traffic directions, $l_k$ is the queue length in direction $k$, $w_k$ is the average waiting time in direction $k$, $o_k$ shows the intersection occupancy status, and $d_{int}$ represents the RV's distance to the intersection.
This observation formulation is consistent with the capabilities of connected vehicles, which can access global traffic states via V2X communication.

We design the reward function to encourage traffic efficiency (reduced average waiting times) and discourage vehicle movement conflicts. When multiple RVs from conflicting streams arrive at the intersection entrance and decide to proceed, priority is assigned based on a combined score of waiting time and queue length.
At time step \( t \), the reward \( r_t \) is defined as \( r_t = \alpha  h_t +  c_t \), where \( h_t \) is the local reward, \(\alpha\) is the weight for local rewards, and \( c_t \) is the conflict penalty. The local reward is given by \( h_t = -w_k \) if Stop, and \( h_t = w_k \) if Go, where \( w_k \) is the average waiting time of all vehicles moving in direction \( k \). The conflict penalty is defined as \( c_t = -1 \) if a conflict occurs, and \( c_t = 0 \) otherwise.
This reward formulation with explicit penalization of conflicts has been shown to reduce unsafe interactions in similar settings~\cite{wang2023learning}.






\vspace{-0.27em}
\subsection{Vehicle Behavior Implementation}
\label{veh_behavior}
The behavior of HVs is governed by Intelligent Driver Model (IDM)~\cite{treiber2013traffic}, which provides acceleration based on surrounding traffic conditions. 
RVs use a hybrid approach: they employ IDM when outside an intersection (30m from the intersection) and execute a learned policy when inside an intersection. 
When the policy indicates forward, the RV accelerates at maximum rate; when signaling to stop, it decelerates based on its current speed $u$ and distance to intersection $d_{int}$: $a = -u^2/2d_{int}$.

\subsection{Complex Intersections and Heterogeneous Traffic}

\begin{figure}[b]
\vspace{-1em} 
    \centering
    \includegraphics[width=0.95\columnwidth]{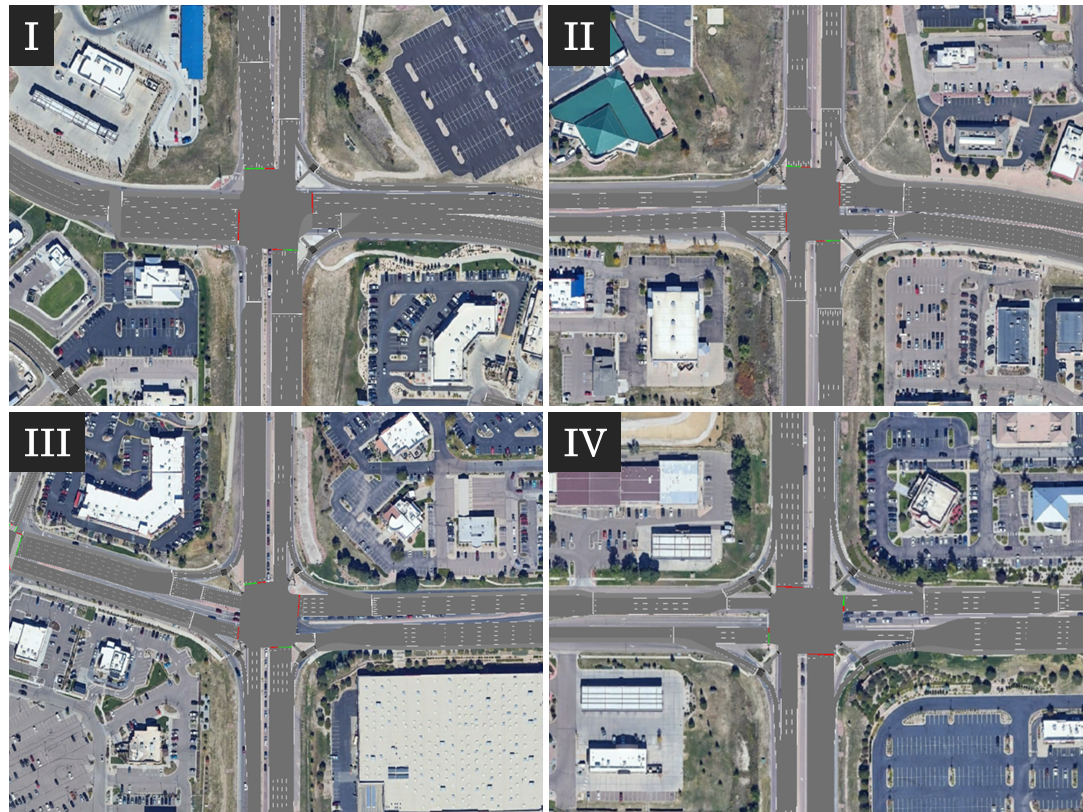} 
    \caption{Real-world and complex intersections used in our study.}
    \label{fig:col_intersections}
\end{figure}

Our study scenarios are four real-world intersections shown in Fig.~\ref{fig:col_intersections}, with the number of incoming lanes 21, 19, 18, and 16 for intersections I, II, III, and IV, respectively. 
Vehicle routes and trajectories are generated using SUMO's routing tools based on turning count data~\cite{wang2023learning}.
To construct heterogeneous traffic, we analyze the real-world dataset I-24 MOTION~\cite{gloudemans202324} and extract the vehicle types and their ratios within: passenger cars (70\%), pickups (3\%), vans (15\%), semi-trailers (11\%), and trucks (1\%)~\cite{poudel2024carl}. 
We ground the vehicle type distribution of our simulation based on these ratios.
The kinematic and dynamic parameters of these vehicle types are presented in Table~\ref{tab:vehicle_properties}~\cite{gloudemans202324, sumo_doc}.


\begin{table}[b]
\caption{The kinematic and dynamic parameters of each vehicle type.}
\vspace{-0.5em}
  \centering
  \scalebox{0.95}{
  \begin{tabular}{|p{2cm}|>{\centering\arraybackslash}p{1.1cm}|>{\centering\arraybackslash}p{1.0cm}|>{\centering\arraybackslash}p{0.6cm}|>{\centering\arraybackslash}p{1.0cm}|>{\centering\arraybackslash}p{0.7cm}|}
    \hline
    \textbf{Parameter} & \textbf{Passenger Car} & \textbf{Pickup} & \textbf{Van} & \textbf{Semi-trailer} & \textbf{Truck} \\
    \hline
    Length (m) & 5   & 5.8 & 5.5 & 16.5 & 7.1 \\
    \hline
    Width (m) & 1.8 & 2   & 2   & 2.55 & 2.4 \\
    \hline
    Height (m) & 1.5 & 1.9 & 2.1 & 4   & 2.4 \\
    \hline
    Mass (kg) & 1500 & 2500 & 3000 & 15000 & 12000 \\
    \hline
    Min Gap~(m) & 2.5 & 2.5 & 2.5 & 2.5 & 2.5 \\
    \hline
    Accel. (m/s$^{2}$) & 2.6 & 2.6 & 2.6 & 1.0 & 1.3 \\
    \hline
    Decel. (m/s$^{2}$) & 4.5 & 4.5 & 4.5 & 4   & 4   \\
    \hline
    Max Speed (m/s) & 55.56 & 33.33 & 27.78 & 36.11 & 36.11 \\
    \hline
  \end{tabular}
  }
  \label{tab:vehicle_properties}
\end{table}


\subsection{Evaluation Metrics}

We provide comprehensive analysis of heterogeneous mixed traffic in terms of traffic efficiency and environmental impact using the following metrics.


\subsubsection{Average Waiting Time}
Average waiting time measures the mean time vehicles spend waiting at the intersection~\cite{wang2023learning}, calculated as \(W_{\text{avg}} = \frac{1}{N} \sum_{i=1}^{N} W_i\), where \(N\) is the total number of vehicles and \(W_i\) is the waiting time of vehicle \(i\). This metric assesses traffic efficiency, with lower values indicating smoother flow and less congestion.

\subsubsection{CO$_2$ Emissions and Fuel Consumption}
We evaluate the environmental impact of heterogeneous mixed traffic using CO$_2$ emissions and fuel consumption, estimated with SUMO's built-in HBEFA3-based model~\cite{sumo_doc3, sumo_doc2}. Each vehicle type is assigned an emission class corresponding to its real-world counterpart (e.g., HBEFA3/PC\_G\_EU4 for passenger cars, HBEFA3/HDV\_D\_EU4 for semi-trailers).
For each vehicle type \( v \), the average CO$_2$ emission is calculated as \( E_{\text{CO}_2}^v = \frac{1}{N_v} \sum_{i=1}^{N_v} E_i \), where \( N_v \) is the total number of vehicles of type \( v \) passing through the intersection, and \( E_i \) is the CO$_2$ emission from vehicle \( i \). We measure fuel consumption in a similar manner. 
Although the HBEFA3-based model does not fully capture the variability of real-world driving, we use it because of its integration with SUMO and its suitability for comparative analysis across scenarios when empirical data is unavailable. 



\subsubsection{Space Headway}
Space headway is measured as the average distance between the front bumper of a vehicle and the front bumper of the vehicle immediately ahead~\cite{treiber2013traffic}: \(H_{\text{space}} = \frac{1}{M} \sum_{j=1}^{M} (X_{j+1} - X_j)\), where \(M\) is the total number of vehicle pairs, \(X_j\) is the position of the vehicle \(j\), and \(X_{j+1}\) is the position of the vehicle immediately ahead of the vehicle \(j\). 
Space headway represents how efficiently vehicles utilize available road space and maintain safe distances.


\section{EXPERIMENTS}
\label{experiments}




We begin by introducing our experiment setup, then present the results on overall traffic efficiency. This is followed by a detailed discussion of both efficiency and environmental impact within heterogeneous mixed traffic. 
\vspace{-0.25em}
\subsection{Experiment Setup}

We model RVs as passenger cars to streamline the complexity of the simulations while still capturing the essence of mixed traffic dynamics. Passenger cars represent the most common vehicle type on urban roads, making them a practical choice for evaluating RV control and coordination. 
Each newly spawned passenger car is randomly designated as an RV or HV based on a pre-specified RV rate.

To evaluate the performance of heterogeneous mixed traffic, we compare two baselines: (1) \textbf{HV-Signalized:} HVs operating under traffic signals, representing the traditional intersection management method; and (2) \textbf{HV-Unsignalized:} HVs without signal control, representing the unsignalized scenarios.
Our study is conducted in a heterogeneous traffic environment, where RVs must navigate the complex dynamics introduced by diverse types of HVs. Consequently, we focus on practical, real-world baselines rather than homogeneous ones, which would fail to capture this core challenge.

For each RV penetration rate (10\% to 90\% in 10\% steps), we train a shared RL policy using Rainbow DQN~\cite{hessel2018rainbow} for 1,000 iterations on an Intel i9-13900KF and NVIDIA RTX 4090, with each run taking 20–30 hours. The network has three layers of 512 units, a learning rate of 0.0005, and a discount factor of 0.99. Each policy is evaluated 10 times over 1,000 seconds. Vehicle type ratios follow Table~\ref{tab:veh_type_ratios}.

\begin{table}[b]
    \vspace{-0.75em} 
    \caption{Vehicle Type Ratios for Different RV Penetration Rates. The initial distribution is extracted from the I-24 MOTION dataset~\cite{gloudemans202324}, with RVs introduced as passenger cars. As RV rate increases, only the passenger car (non-RV) ratio decreases until 70\%, after which other vehicle type ratios are proportionally reduced to maintain the total at 100\%.}
    \begin{adjustbox}{max width=\columnwidth}
    \begin{tabular}{|>{\centering\arraybackslash}p{0.6cm}|>{\centering\arraybackslash}p{1.8cm}|>{\centering\arraybackslash}p{0.7cm}|>{\centering\arraybackslash}p{0.5cm}|>{\centering\arraybackslash}p{1.4cm}|>{\centering\arraybackslash}p{0.7cm}|}
        \hline
        \textbf{RV (\%)} & \textbf{Passenger Car (non-RV) (\%)} & \textbf{Pickup (\%)} & \textbf{Van (\%)} & \textbf{Semi-trailer (\%)} & \textbf{Truck (\%)} \\ \hline
        10                    & 60                                     & 3                    & 15                & 11                 & 1                   \\ \hline
        20                    & 50                                     & 3                    & 15                & 11                 & 1                   \\ \hline
        30                    & 40                                     & 3                    & 15                & 11                 & 1                   \\ \hline
        40                    & 30                                     & 3                    & 15                & 11                 & 1                   \\ \hline
        50                    & 20                                     & 3                    & 15                & 11                 & 1                   \\ \hline
        60                    & 10                                     & 3                    & 15                & 11                 & 1
        \\ \hline
        70                    & 0                                      & 3                    & 15                & 11                 & 1
        \\ \hline
        80                    & 0                                      & 2                    & 10                & 7.3               & 0.7                \\ \hline
        90                    & 0                                      & 1                    & 5                 & 3.7               & 0.3                \\ \hline
    \end{tabular}
    \end{adjustbox}
    \label{tab:veh_type_ratios}
\end{table}


\subsection{Overall Traffic Efficiency}

\begin{figure}[b]
    \centering
    \vspace{-0.85em}
    \includegraphics[width=0.98\columnwidth]{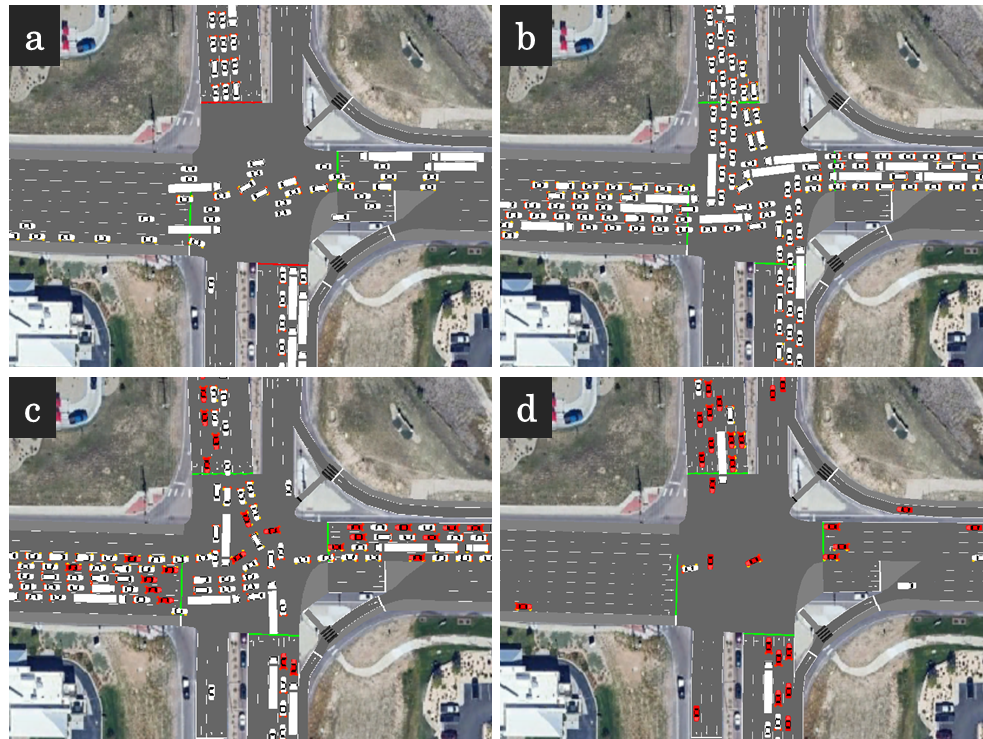} 
    \vspace{-0.5em}
    \caption{Traffic at intersection I at the 10-minute mark (RVs in red, HVs in white). (a) HV-Signalized: Flow is controlled by traffic lights. (b) HV-Unsignalized: Severe gridlock emerges without signals. (c) 10\% RV: Slight improvement, but HV dominance limits coordination. (d) 60\% RV: RVs adaptively coordinate, reducing congestion and enhancing flow.}

    \label{fig:sim_demo}
\end{figure}

The traffic demands for the four unsignalized intersections are extracted from real-world data~\cite{wang2023learning}. Specifically, intersection I has 1,157 vehicles per hour (v/h), intersection II has 1,089 v/h, intersection III sees 928 v/h, and intersection IV experiences the lowest demand at 789 v/h. 
Table~\ref{tab:awt_intersections}  highlights the results for average waiting time at four intersections under different scenarios.  
Across all four intersections, increasing the RV rate leads to substantial reductions in average waiting times compared to both baseline scenarios, indicating improved traffic efficiency. 
At intersection I, the waiting time decreases from 122.35 s at 10\% RV to 12.90 s at 90\% RV, achieving maximum improvements of \textbf{65.48\% and 89.92\% over HV-Signalized and HV-Unsignalized}. At intersection II, the highest improvement comes at RV rate 80\% (8.32 s), providing a \textbf{86.39\% gain over HV-Signalized and 67.42\% over HV-Unsignalized baselines}. 
Intersections III and IV also demonstrate significant gains, with \textbf{maximum improvements of 50.13\% and 77.54\% over signalized, and 77.65\% and 91.19\% over unsignalized baselines}, respectively. These results highlight the potential of RVs in 
reducing congestion and improving throughput of mixed traffic, compared to traditional control method via traffic lights. Fig.~\ref{fig:sim_demo} shows an example scenario at intersection I.

\begin{table*}[t]
\caption{Average Waiting Time (s) at four intersections under different traffic control scenarios. Higher RV rates significantly reduce waiting times across all intersections compared to HV-Signalized and HV-Unsignalized, especially at 60\% RV rate or above. For Intersection I, waiting time drops from 122.35s at 10\% RV rate to 12.90s at 90\%. Intersection II performs best at 80\% RV rate, reducing waiting time by 86.39\%. Intersections III and IV show variable improvements but still achieve substantial reductions at higher RV rates (e.g., 60\% and 50\%, respectively).}

\centering
\begin{tabularx}{\textwidth}{@{}l p{1.8cm} p{2.2cm} XXXXXXXXXX@{}}
\toprule
\textbf{Intersection} & \textbf{HV-Signalized} & \textbf{HV-Unsignalized} & \multicolumn{9}{c}{\textbf{RV Penetration Rate}} \\ \cmidrule(l){4-12} 
 & & & \textbf{10\%} & \textbf{20\%} & \textbf{30\%} & \textbf{40\%} & \textbf{50\%} & \textbf{60\%} & \textbf{70\%} & \textbf{80\%} & \textbf{90\%} \\ \midrule
\textbf{I}  & 37.37 & 127.97 & 122.35 & 97.16 & 90.85 & 47.78 & 38.52 & 21.70 & 23.31 & 17.31 & \textbf{12.90} \\ \midrule
\textbf{II} & 61.13 & 25.54  & 31.88  & 38.13 & 11.72 & 16.77 & 9.54  & 8.34  & 9.63  & \textbf{8.32}  & 8.81  \\ \midrule
\textbf{III} & 50.61 & 112.94 & 105.07 & 81.71 & 62.95 & 46.51 & 50.15 & \textbf{25.24} & 39.94 & 44.46 & 31.05 \\ \midrule
\textbf{IV} & 39.71 & 101.32 & 54.08  & 56.33 & 50.12 & 14.59 & \textbf{8.92}  & 22.11 & 17.87 & 11.83 & 10.39 \\ 
\bottomrule
\end{tabularx}
\label{tab:awt_intersections}
\vspace{-0.75em} 
\end{table*}


\subsection{Traffic Heterogeneity Evaluation}


The integration of RVs into traffic systems presents challenges due to the diverse vehicle types on urban roads. 
Differences in size, speed, and maneuverability complicate coordination, particularly at intersections. 
This study analyzes how various vehicle types---passenger cars, pickups, vans, semi-trailers, and trucks---respond to rising RV rates in terms of traffic efficiency and environmental impact.

\subsubsection{Average Waiting Time}
\begin{figure*}[t]
    \centering
    \includegraphics[width=1.0\textwidth]{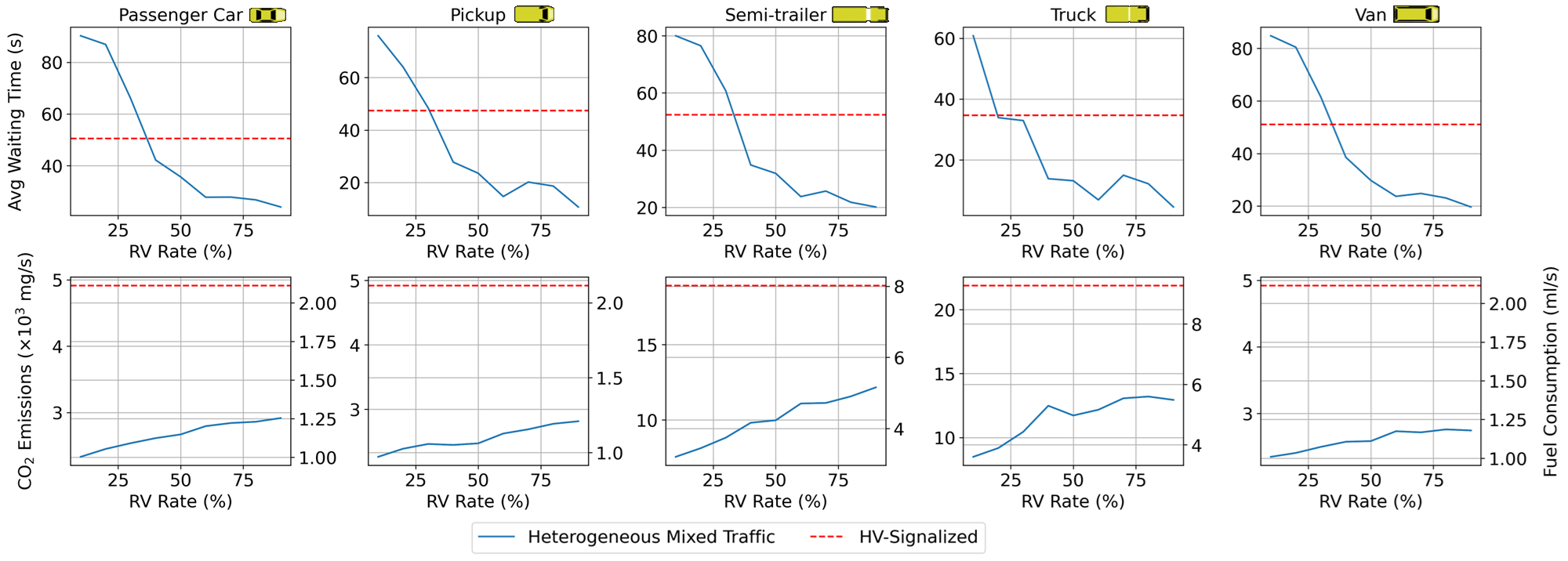}
    \vspace{-1.75em} 
    \caption{Waiting Time (top row), CO$_2$ Emissions and Fuel Consumption (bottom row) for different vehicle types across RV rates (10\% to 90\%) compared to HV-Signalized, averaged over a 1000-second run. As RV rate increases, waiting times generally decrease for all vehicle types, outperforming HV-Signalized starting from 40\% RV penetration. Due to their near-identical trends in SUMO’s emission model, a single line is used to represent both CO$_2$ emissions and fuel consumption in the bottom row. These metrics slightly increase but stay below HV-Signalized levels. Larger vehicles, like trucks and semi-trailers, show more pronounced increases due to size and weight, while smaller vehicles benefit from reduced waiting times with moderate environmental impact.}

    \label{fig:awt_fc_speed_vehtype}
\vspace{-1.0em} 
\end{figure*}

Fig.~\ref{fig:awt_fc_speed_vehtype} (top row) presents the average waiting times at intersections for different vehicle types across RV rates, compared to HV-Signalized. 
As the RV rate approaches 0\% in our unsignalized intersections, the results align with the HV-Unsignalized baseline. This explains why the waiting times at 10\% RV rate are higher than HV-Signalized baseline for all vehicle types. As demonstrated in our overall results, the HV-Unsignalized scenario performs significantly worse than HV-Signalized on average.  
We observe that all vehicle types experience a decrease in waiting time as RV rate increases, indicating a broad positive impact of RV introduction on traffic efficiency. However, the magnitude of improvement varies across vehicle types. 

Trucks exhibit the most significant improvement, with waiting time reducing from 60.80 s at 10\% RV rate to 4.50 s at 90\% RV rate---\textbf{an 87.03\% reduction compared to HV-Signalized (34.68 s)}. 
Pickups, with the second lowest percentage in the traffic, show similar patterns of improvement---\textbf{up to 77.52\% reduction vs. HV-Signalized}. Semi-trailers show less dramatic improvements, with waiting time reducing from 79.97 s to 20.12 s---\textbf{61.57\% reduction compared to  HV-Signalized}. 
Passenger cars and vans constitute the majority of traffic, showing significant but comparatively smaller improvements. Passenger cars see their average waiting time decrease to 23.99 s, while vans improve to 19.68 s, \textbf{achieving 52.46\% and 61.51\% improvements}, respectively, compared to HV-Signalized. 
The results show varied impacts of RV coordination, with less frequent vehicles like trucks and pickups experiencing the greatest reductions in waiting times, while more common vehicles like passenger cars and vans show moderate improvements.

\begin{table}[t]


\caption{Percentage improvements in waiting times for vehicle types compared to HV-Signalized. Bold values indicate highest improvement for each vehicle type at 90\% RV rate, demonstrating rarity advantage for less common vehicles 
and benefits of reduced heterogeneity for passenger cars. 
}


\label{tab:awt_percentage_reduction_vehtype}
\small 
\begin{tabular}{@{}lrrrrrr@{}}
\toprule
\multirow{2}{*}{\textbf{Vehicle Type}} & \multicolumn{6}{c}{\textbf{RV Penetration Rate}} \\
\cmidrule(l){2-7}
 & \textbf{40\%} & \textbf{50\%} & \textbf{60\%} & \textbf{70\%} & \textbf{80\%} & \textbf{90\%} \\
\midrule
Passenger Car & 16.49 & 29.40 & 45.00 & 44.89 & 46.98 & \textbf{52.46} \\
Pickup    & 41.40 & 50.36 & 69.09 & 57.46 & 60.67 & \textbf{77.52} \\
Semi-trailer      & 33.55 & 39.16 & 54.63 & 50.95 & 58.41 & \textbf{61.57} \\
Truck     & 60.22 & 62.14 & 80.30 & 56.88 & 65.02 & \textbf{87.03} \\
Van       & 24.58 & 41.78 & 53.58 & 51.43 & 54.82 & \textbf{61.51} \\
\bottomrule
\end{tabular}
\vspace{-1.75em} 
\end{table}

To better understand this trend, Table~\ref{tab:awt_percentage_reduction_vehtype} shows percentage improvements in waiting times for each vehicle type at key RV rates compared to HV-Signalized.
The observed ``rarity advantage,'' where less frequent vehicles like trucks and pickups show greater improvements than common vehicles, can be attributed to temporal spacing, i.e., their ability to take advantage of the gaps created by RV coordination without contributing significantly to overall congestion. 
Their infrequent presence at intersections means they are more likely to encounter optimized traffic conditions created by the RVs. 
Similarly, semi-trailers' moderate improvement compared to trucks might be due to their higher frequency in traffic, leading to more vehicle interactions and potentially more complex maneuvering requirements. In contrast, passenger cars (70\% of traffic) frequently encounter other vehicles competing for intersection access. 
Interestingly, we observe a slight increase in waiting times for most vehicle types when the RV rate reaches 60--70\%, before decreasing again at higher rates. This non-monotonic pattern suggests a potential phase transition of coordination complexity within traffic dynamics, as the system shifts from HV-dominated to RV-dominated. 
At higher RV rates (80--90\%), the greater uniformity in vehicle control strategies allows for more efficient system-wide coordination, resulting in the resumed decrease in waiting times. 
The dip may also stem from a cascading effect in the sequential intersection layout, where efficient RV flow from upstream intersections leads to dense platoons that challenge coordination downstream.
This phenomenon shows the complex, non-linear relationship between RV rates and traffic efficiency under heterogeneous settings.

\subsubsection{CO$_2$ Emissions and Fuel Consumption}


Fig.~\ref{fig:awt_fc_speed_vehtype} (bottom row) shows CO$_2$ emissions and fuel consumption trends across vehicle types as RV rate increases.
For CO$_2$ emissions, we observe a general increasing trend for all vehicle types as RV rates rise. Passenger cars show a moderate increase, with emissions rising from 2,333 mg/s at 10\% RV rate to 2,917 mg/s at 90\% RV rate, a 25\% increase. Pickups exhibit a similar trend, experiencing a 24\% rise.
Larger vehicles demonstrate more substantial increases in CO$_2$ emissions. Trucks, For example, trucks see their emissions rise from 8,497 mg/s at 10\% RV rate to 12,946 mg/s at 90\% RV rate, leading to a 52\% increase. Semi-trailers show the most dramatic change, with emissions going from  7,554 mg/s to 12,165 mg/s, a significant 61\% increase. 
The overall trend highlights that CO$_2$ emissions rise more sharply for larger vehicles as RV rate increases. 
Fuel consumption patterns closely mirror the CO$_2$ emission trends.
Semi-trailers show the most significant increase, with fuel consumption rising from about 3.2 ml/s at 10\% RV rate to 5.2 ml/s at 90\% RV rate, a 62.5\% increase. Trucks follow a similar pattern, with consumption increasing from 3.6 ml/s to 5.6 ml/s, a 55.5\% rise. Passenger cars, pickups, and vans exhibit lower overall fuel consumption but still show an increasing trend with higher RV rates. For instance, passenger car fuel consumption rises from 1.0 ml/s to 1.25 ml/s, a 25\% increase as RV rate goes from 10\% to 90\%. 
Overall, larger vehicles experience the greatest rise in fuel consumption at higher RV rates, while smaller vehicles see more moderate increases. 


The rise in both CO$_2$ emissions and fuel consumption for all vehicle types could be attributed to the shift from intermittent idling to more continuous motion. While vehicles spend less time stationary, they travel at higher speeds, potentially increasing overall fuel consumption and emissions. Due to their weight and lower fuel efficiency at higher speeds, larger vehicles like trucks and semi-trailers are more sensitive to this change, resulting in a net increases in  emissions and fuel consumption. 
However, both CO$_2$ emissions and fuel consumption remain significantly lower than HV-Signalized across all vehicle types and RV rates. This suggests that while there is a trade-off between reduced waiting times and increased emission/fuel consumption as RV rate rises, the net impact of RV-managed intersections remains more environment-friendly than signalized intersections.

\subsubsection{Space Headway} 

\begin{figure}[t]
    \centering
    \includegraphics[width=0.98\columnwidth]{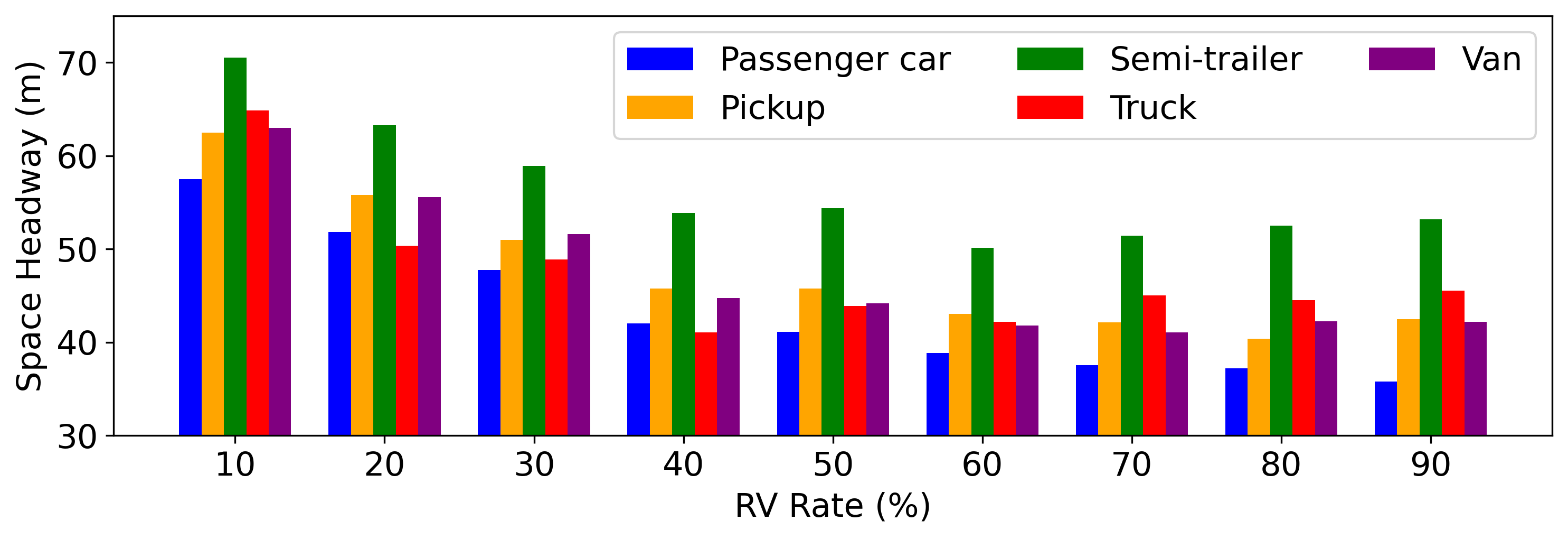}
    \vspace{-1.0em}
    \caption{ Average space headways for different vehicle types at varying RV penetration rates. As RV rate increases, space headways decrease across all vehicle types, leading to better road space utilization. Passenger cars show the largest reduction. Trucks and semi-trailers see smaller reductions, with a plateau effect at higher RV rates.}
    \label{fig:headway_comparison} 
\vspace{-1.5em} 
\end{figure}

As shown in Fig.~\ref{fig:headway_comparison}, we observe a consistent reduction in space headways across all vehicle types as RV rate increases.  
The impact of RVs on space headways varies across vehicle types. Passenger cars, which dominate the traffic composition, show a marked decrease in space headways from 57.49 m at 10\% RV rate to 35.78 m at 90\% RV rate, a reduction of about 38\%. Larger vehicles such as semi-trailers, which typically require more space due to their size and slower acceleration, also exhibit reduced space headways, with space headways decreasing from 70.55 m to 53.21 m (25\% reduction) as RV rate increases from 10\% to 90\%. 
These results underscore the potential of RVs in facilitating streamlined traffic flows. 
In case of less occurring vehicles (e.g., trucks and pickups) at higher RV rates, we observe a plateau effect, where headways stabilize or slightly increase. This could be attributed to the fact that low-frequency vehicles interact less frequently at intersections and face fewer congestion issues in high-RV scenarios. Therefore, they do not experience the same pressure to reduce headways as passenger cars do. The changing vehicle type mix at higher RV rates may also contribute to this effect.

\section{CONCLUSION AND FUTURE WORK}
\label{conclusion}

This study investigates heterogeneous mixed traffic control using RVs in complex environments. Results show that higher RV rates significantly reduce average waiting times for all vehicle types—up to 86\% and 91\% lower than HVs at signalized and unsignalized intersections, respectively. The rarity advantage suggests certain vehicle types benefit more from RV control, indicating that adaptive strategies with vehicle-type awareness could further boost efficiency. While RVs raise emissions and fuel use due to higher speeds, overall efficiency still surpasses signalized baselines. We also observe reduced space headways across all vehicle types, with the largest drop for passenger cars.


There exist several future directions.  
First, testing RV operation in more unseen environments---such as roundabouts, three-way intersections, and large-scale networks---and with imperfect information from local sensors or lossy V2V communication would better assess policy robustness and generalization.  
Second, adopting a continuous action space could enable more efficient coordination than the current Stop/Go model.  
Third, this study uses only passenger cars as RVs for tractability; extending to other vehicle types may allow vehicle-specific reward shaping (e.g., fuel-efficient truck behaviors).  
Finally, broader comparisons with alternative RL or optimization-based intersection management strategies remain an important future direction.

\section*{Acknowledgments}
This research is supported by NSF 2153426, 2332210, 2524239, and 2129003. The authors also thank NVIDIA and the Tickle College of Engineering at University of Tennessee, Knoxville for their support.









\bibliographystyle{./IEEEtran} 
\bibliography{./references} 

\end{document}